\documentclass[apl, reprint, twocolumn]{revtex4}
\usepackage[utf8]{inputenc}
\usepackage{graphicx}% Include figure files
\usepackage{dcolumn}% Align table columns on decimal point
\usepackage{bm}% bold math
\usepackage{soul} % to cross out usin \textst{}
\usepackage{color} % to use different text colours using \textcolor{red}{}

\begin{document}
\title{Phase segregation in Mg$_{x}$Zn$_{1-x}$O probed by optical absorption and photoluminescence at high pressure}
\author{V. Mar\'in-Borr\'as}
\author{J. Ruiz-Fuertes}
  \email{javier.ruiz-fuertes@uv.es}
\author{A. Segura}
\author{V. Mu\~{n}oz-Sanjos\'e}
\affiliation {Departament de F\'isica Aplicada, Universitat de Val\`encia, Dr. Moliner 50, 46100 Burjassot, Spain}
\date{\today}% It is always \today, but any date may be explicitly specified

\begin{abstract}
The appearance of segregated wurtzite Mg$_x$Zn$_{1-x}$O with low Mg content in thin films with $x>0.3$ affected by phase separation, cannot be reliably probed with crystallographic techniques owing to its embedded nanocrystalline configuration. Here we show a high-pressure approach which exploits the distinctive behaviors under pressure of wurtzite Mg$_x$Zn$_{1-x}$O thin films with different Mg contents to unveil phase segregation for $x>0.3$. By using ambient conditions photoluminescence (PL), and with optical absorption and PL under high pressure for $x=0.3$ we show that the appearance of a segregated wurtzite phase with a magnesium content of x $\sim$ 0.1 is inherent to the wurtzite and rock-salt phase separation. We also show that the presence of segregated wurtzite phase in oversaturated thin films phase is responsible for the low-energy absorption tail observed above $x=0.3$ in our Mg$_x$Zn$_{1-x}$O thin films. Our study has also allowed us to extend the concentration dependence of the pressure coefficient of the band gap from the previous limit of $x$ = 0.13 to $x \approx 0.3$ obtaining d$E_g$/d$P$ = 29 meV/GPa for wurtzite with $x \approx 0.3$ and 25 meV/GPa for the segregated $x\approx 0.09$ wurtzite phase.

%Phase segregation of Mg$_x$Zn$_{1-x}$O into Mg$_{0.09}$Zn$_{0.91}$O for Mg contents above $x=0.3$ has been shown with ambient conditions photoluminescence (PL) for $x>0.35$ and unveiled for $x=0.3$ with optical absorption and PL under high pressure in spray-pyrolysis as grown thin films. Our results show that the appearance of a segregated portion of wurtzite phase with an equilibrium Mg content of $x = 0.09$ is inherent to the wurtzite and rock-salt phase separation always observed above $x=0.3$ in Mg$_x$Zn$_{1-x}$O thin films. The value found in this work on as grown thin films is compatible with the equilibrium Mg content of $x=0.15$ found by other authors on thermal annealed thin films grown by molecular beam epitaxy. The pressure coefficient of the band gap d$E_g$/d$P$ = 29 meV/GPa obtained for Mg$_{0.3}$Zn$_{0.7}$O is clearly smaller than the previously estimated of 36.6 meV/GPa. This indicates that for high Mg concentrations the effect of compression on the band gap of Mg$_x$Zn$_{1-x}$O saturates.
\end{abstract}

\maketitle

\section{Introduction}
With a deep excitonic binding energy of 59 meV \citep{mang1995}, long-lived optical phonons \citep{millot2010}, and a polar structure, zinc oxide (ZnO) has been extensively studied among others applications for stimulated light emission \citep{tang1997,wille2016}, as a second harmonic generation based O$_2$ sensor \citep{andersen2014}, and as a matrix for diluted magnetic semiconductors \citep{gilliland2012}. Alloying with magnesium its band gap widens from 3.37 eV to 7.8 eV \citep{ohtomo1998,kumar2013,schleife2011,pantelides1974} reinforcing its use in solar-blind communication devices \citep{liu2009}. ZnO crystallizes in the hexagonal wurtzite-type structure while MgO crystallizes in the cubic rock-salt-type structure. Therefore, once the solubility limit is reached in wurtzite-type Mg$_x$Zn$_{1-x}$O solid solution, $\sim$ 4\% in bulk \citep{segnit1965} and $\sim$ 30\% or $\sim$ 50\% in thin films depending on the growth method \citep{kumar2013,redondo2012}, phase separation appears \citep{gries2015} and both rock-salt and wurtzite type phases coexist. A lot of effort has been put on trying to reach the highest incorporation limit of Mg$^{2+}$ in phase pure wurzite-type Mg$_x$Zn$_{1-x}$O. However, the study of the optical properties of Mg$_x$Zn$_{1-x}$O thin films with phase separation has been scarce \citep{thapa2013,huso2014,lopez2015}. Those previous studies find with optical absorption spectroscopy \citep{thapa2013,lopez2015} a low-energy absorption tail overlapped to the main absorption edge for Mg concentrations above a critical concentration when phase separation occurs. The origin of this absorption tail observed in Mg$_x$Zn$_{1-x}$O thin films grown by different methods, has been tentatively explained by some authors \citep{lopez2015} as due to the beginning of the absorption edge of the coexisting rock-salt phase. However, the rock-salt MgO even alloyed with Zn remains transparent up to energies quite above 4.06 eV \citep{segura2003}. Therefore, the origin of the observed low-energy absorption tail remains unclear. The study of \citet{gries2015} has successfully employed transmission electron microscopy (TEM) on thermally annealed Mg$_{0.3}$Zn$_{0.7}$O thin films grown by molecular beam epitaxy (MBE) to investigate the microscopic effect of having phase separation. They found that the coexistence of the wurtzite and rock-salt phases in Mg$_x$Zn$_{1-x}$O due to phase separation gives rise to the existence of an secondary wurtzite-type phase, with a reduced Mg content of $x \approx 0.15$ determined by TEM energy dispersive x-ray spectroscopy (EDX). The presence of some amount of this segregated wurtzite-type phase could explain the low-energy absorption tail observed by optical absorption studies on Mg$_x$Zn$_{1-x}$O thin films when phase separation occurs. However, neither x-ray diffraction (XRD) nor TEM are well suited for probing phase segregation in as grown Mg$_x$Zn$_{1-x}$O thin films with phase separation. The volume of this segretated wutzite phase was too low to be detected by XRD and it was embedded in the thin film preventing selective access with TEM. For this reason,  \citet{gries2015} had to employ a buffer layer of MgO/ZnO to promote the growth of a segregated wurtzite phase in the Mg$_{x}$Zn$_{1-x}$O thin film, after annealing the sample at 950 $^{\circ}$C, in order to detect the segregated wurtzite phase by means of TEM.   

Here we show a spectroscopic approach to probe phase segregation in spray pyrolysis (SP) as grown Mg$_{0.3}$Zn$_{0.7}$O thin films by means of optical absorption spectroscopy and PL measurements under high pressure and ambient temperature along the composition-dependent pressure-induced wurtzite to rock-salt irreversible phase transition \citep{decremps2002,sans2004,desgreniers1998} avoiding any postgrowth treatment on the grown sample.

\begin{figure}
\centering
\includegraphics[width=0.45\textwidth]{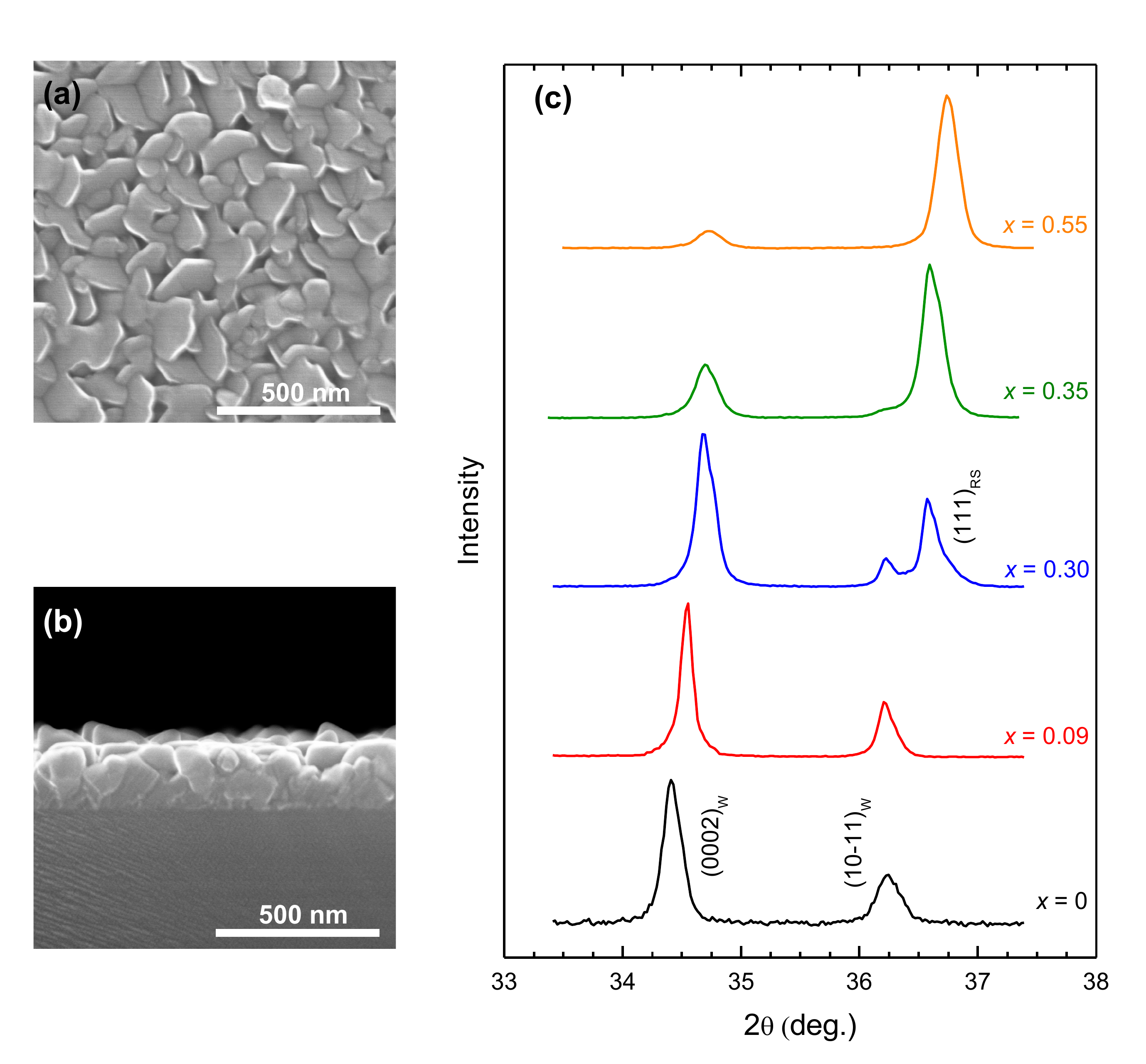}
\caption{\label{fig:fig1} Scanning electron microscope (SEM) images of top (a) and cross-sectional (b) views of as grown 250-nm thick Mg$_{0.3}$Zn$_{0.7}$O thin film on a $c$-oriented sapphire substrate. (c) Normalized x-ray diffractograms of Mg$_{x}$Zn$_{1-x}$O thin films of different Mg contents $x$. The hexagonal (0002)$_w$ and (10$\overline{1}$1)$_w$ reflections of the wurtzite as well as the cubic (111)$_{rs}$ reflection of rock-salt are indicated. The labels shown next to each diffractogram are the Mg concentrations measured in the sample.}
\end{figure} 

\section{Experimental Details}
Thin films of Mg$_x$Zn$_{1-x}$O with measured Mg content of $x$ = 0, 0.06, 0.09, 0.15, 0.22, 0.3, 0.35, and 0.55 were grown by the SP method on $c$-plane oriented sapphire, as explained in the work of \citet{lopez2015}, and on $c$-oriented ScAlMgO$_4$ substrates in the case of $x$ = 0.3 for studies under high pressure. The samples were characterized by scanning electron microscopy (SEM) to see their morphology and by XRD. The thicknesses of the thin films were between 150 and 500 nm depending on the Mg content. XRD diffractograms were measured using a Bruker D8 Advance A25 diffractometer with Cu K$\alpha_1$ wavelength. For the optical absorption and the PL measurements we used a deuterium lamp and an all-solid-state pulsed laser at 266 nm with a maximum power of 10 mW, respectively. The transmitted or photoemitted light were detected with a multichanel UV-enhanced spectrometer. For the high-pressure experiments a confocal system with two cassegrain objectives was employed together with the same UV-Vis spectrometer. In the high-pressure experiments we used the Mg$_{0.3}$Zn$_{0.7}$O sample grown on the ScAlMgO$_4$ substrate which was exfoliated to a thickness of around 10 $\mu$m. ScAlMgO$_4$ has been shown to have the same compressibility \citep{errandonea2011,desgreniers1998} as ZnO. The sample was loaded in a diamond anvil cell (DAC) equipped with two diamonds with 500 $\mu$m culets, in the center of a 250 $\mu$m hole made in an Inconel gasket preindented to a thickness of 45 $\mu$m. Inside the pressure chamber we placed a ruby chip for pressure calibration \citep{mao1978} and a mixture of methanol-ethanol (4:1) as pressure transmitting medium. 

\section{Results and Discussion}

In Fig. \ref{fig:fig1} (a) and (b) we show the SEM images of the top and cross-sectional views of the Mg$_{0.3}$Zn$_{0.7}$O thin film grown on the sapphire substrate. One can appreciate that the sample presents a typical uniform and embedded leaves characteristic of ZnO grown by SP. This morphology, only shown for the $x=0.3$ sample for clarity, stays constant only evolving into a grain-like shape for higher Mg contents. As expected, Mg incorporation results into a shift of the peak position corresponding to the hexagonal (0002) reflection towards higher $2\theta$ angles as the result of the contraction of the $c$ lattice parameter observed before \citep{lopez2015,kumar2013}. Regarding the (10$\overline{1}$1) reflection peak position, since it is contributed by both lattice parameters and $a$ expands differently to $c$ with Mg incorporation, it remains almost unaffected by Mg incorporation. Above $x= 0.3$ the peak corresponding to the (0002) reflection broadens as an indicative of compositional disorder and the cubic rock-salt reflection (111) peak emerges and grows with Mg incorporation. This confirms that our samples are phase pure wurtzite up to $x= 0.3$ when the onset of the phase separation occurs. The presence of spinel (Mg,Zn)Al$_2$O$_4$ known to appear as a spurious phase in some processes of synthesis was not detected neither by TEM \citep{lopez2015} nor XRD.

\begin{figure}
\centering
\includegraphics[width=0.45\textwidth]{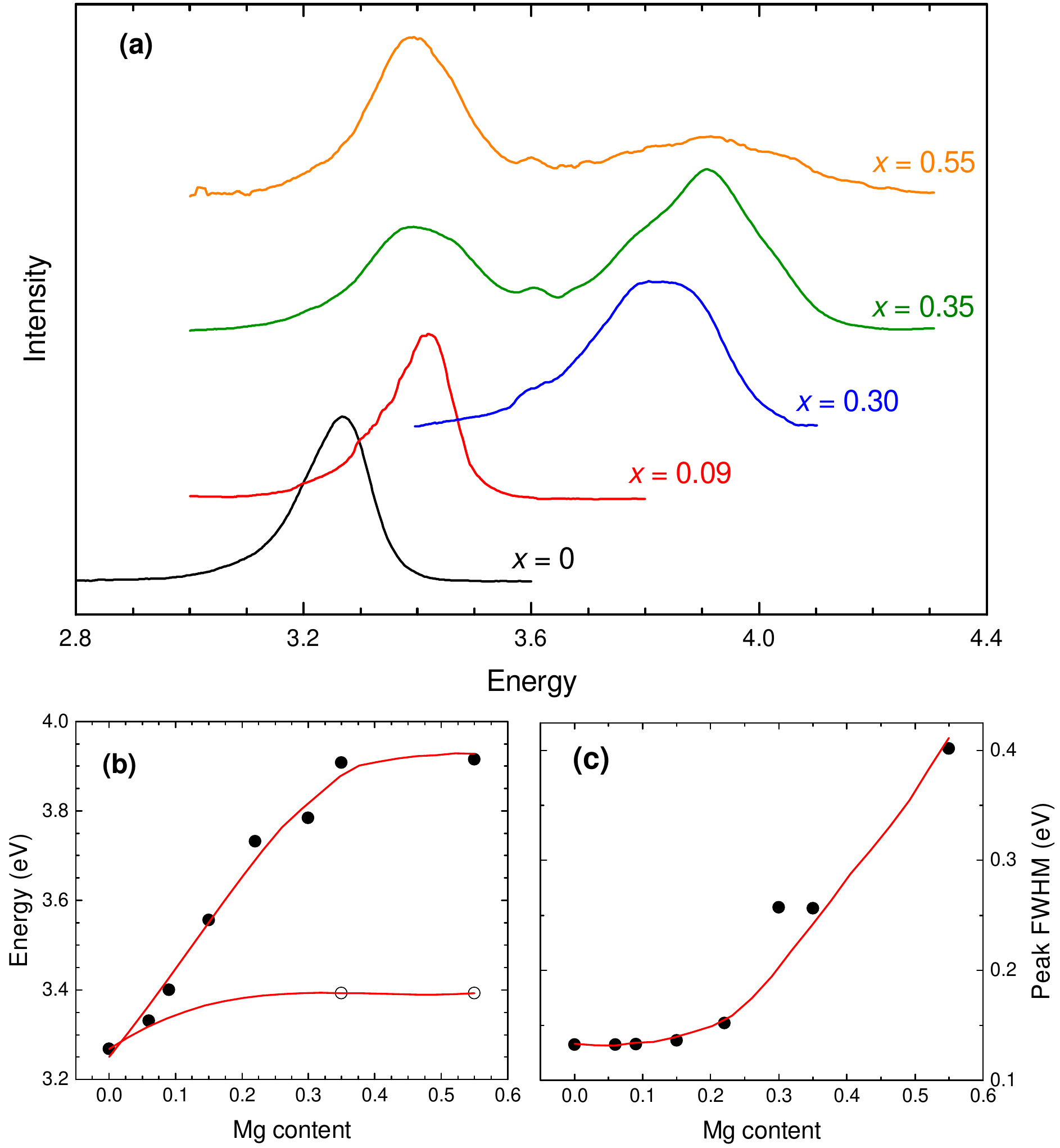}
\caption{\label{fig:fig2} (a) Photoluminescence (PL) spectra of Mg$_{x}$Zn$_{1-x}$O of some of our thin films with different Mg content $x$. (b) Dependence of the maximum of the PL peak and (c) FWHM of the PL peak with Mg content obtained from the fit to an asymmetric lorentzian function. Continuous lines are guides for the eye. The empty dots are the maximum of the extra PL peak that emerges at lower energy for $x >$ 0.3 eV. All measurements were performed with the same integration time in transmittance mode.}
\end{figure}

\begin{figure*}
\centering
\includegraphics[width=0.9\textwidth]{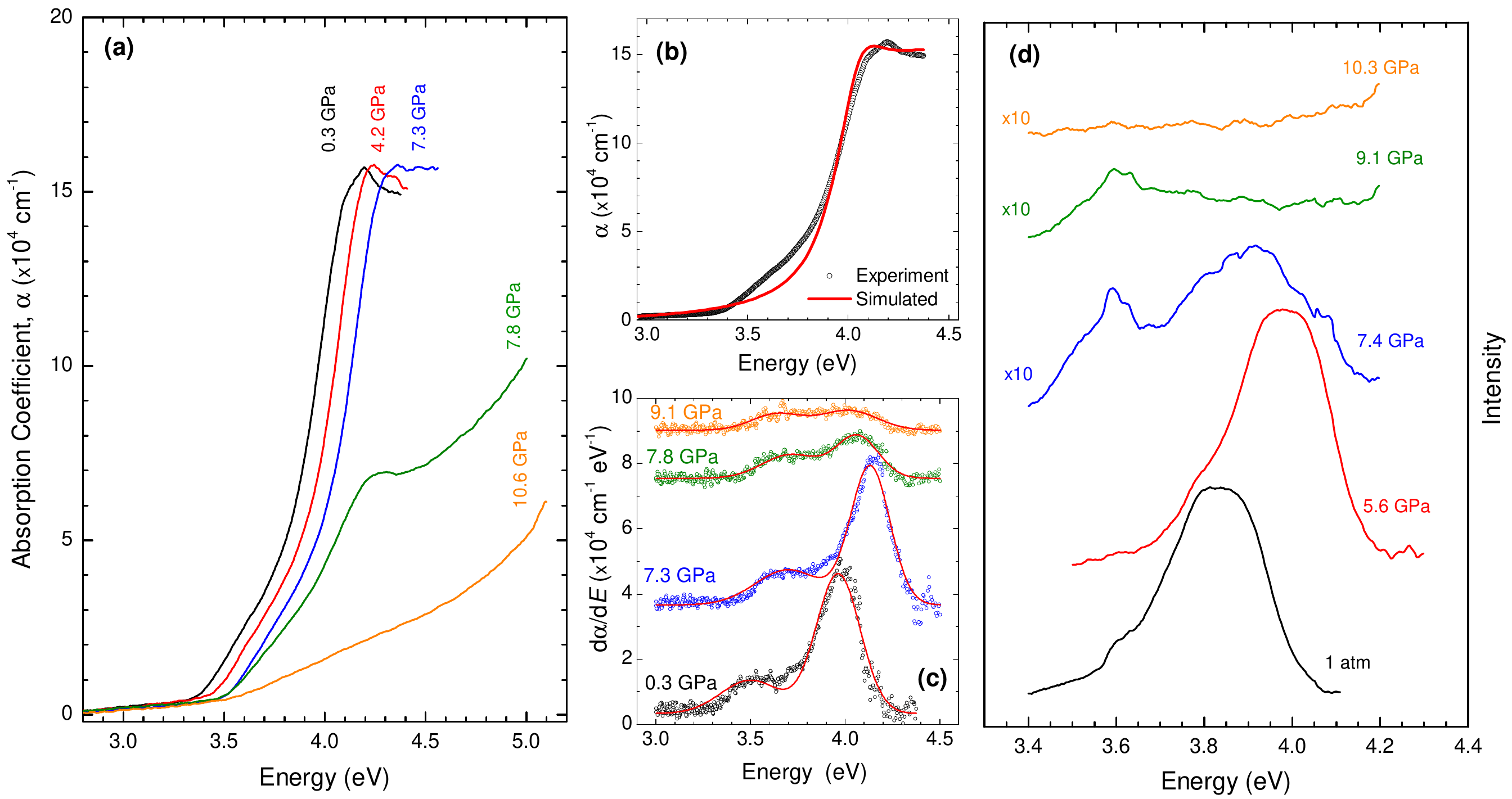}
\caption{\label{fig:fig3} (a) Absorption edge of the Mg$_{0.3}$Zn$_{0.7}$O sample deposited on ScAlMgO$_4$ at different pressures. (b) Simulated absorption spectrum (continuous line) of Mg$_{0.3}$Zn$_{0.7}$O at 0.3 GPa according to Elliot-Toyozawa theory \citep{elliot1957,toyozawa1958,goni1990} together with the experimental spectrum (black dots). (c) Energy derivative of the absorption spectra at different pressures. Red continuous lines show the fit to two gaussians while the dots are the experimental data. (d) PL of the Mg$_{0.3}$Zn$_{0.7}$O sample deposited on ScAlMgO$_4$ at different pressures.}
\end{figure*}

The PL spectra of Mg$_{x}$Zn$_{1-x}$O thin films with different measured Mg content are shown in Fig. \ref{fig:fig2} together with the dependence of the peak maximum and the full width at half maximum (FWHM) with Mg concentration. As expected, the PL peak of wurtzite-type Mg$_{x}$Zn$_{1-x}$O blueshifts from 3.27 eV for $x=0$ to 3.91 eV for $x=0.35$ [Fig. \ref{fig:fig2} (b)]. Also, from $x=0.3$, under Mg incorporation the FWHM of the main PL peak [Fig. \ref{fig:fig2} (c)] starts to broaden due to the disorder caused by the presence of the phase separation found with XRD [Fig. \ref{fig:fig1} (c)]. According to \citet{gries2015} the presence of phase separation in Mg$_{x}$Zn$_{1-x}$O would give rise to the appearance of segregated wurtzite phase with less Mg content. This would result into the appearance of a second PL peak at lower energies. We do not observe the additional PL peak for $x=0.3$, when phase separation starts in our samples, but above $x=0.3$ an additional peak emerges at $\sim$3.4 eV. An energy that approximately corresponds to the PL of a wurtzite-type Mg$_{x}$Zn$_{1-x}$O sample with measured Mg content of $x=0.09$ [Fig. \ref{fig:fig2} (a)]. Similarly to \citet{gries2015} we do not find within our resolution any energy change of the PL peak at 3.4 eV with Mg content indicating that in our thin films the equilibrium Mg concentration of the segretated wurtzite phase is $x \approx 0.09$. However, the question that arises is why we only find the PL peak at $\sim$3.4 eV above $x=0.3$ if phase separation already starts at $x=0.3$ in our as grown thin films according to XRD [Fig. \ref{fig:fig1} (c)]. We shall address this issue below. 

As commented before, previous optical absorption spectroscopy studies \citep{thapa2013,lopez2015} on as grown wurtzite-type Mg$_{0.3}$Zn$_{0.7}$O thin films show that for this Mg concentration a low-energy absorption tail appears overlapping to the main absorption edge of the sample. In Fig. \ref{fig:fig3} (a) we show the absorption edge of our wurtzite-type Mg$_{0.3}$Zn$_{0.7}$O thin film as grown on $c$-oriented ScAlMgO$_4$ with a thickness of 150 nm. Although the presence of the excitonic absorption indicates the high crystalline quality of the sample, the low-energy absorption tail is clear when we compare the experimental data with the calculated absorption edge according to the Elliot-Toyozawa's theory \citep{elliot1957,toyozawa1958,goni1990} considering a single absorption edge [Fig. \ref{fig:fig3} (b)]. Since the rock-salt phase is transparent in this energy range, the low-energy absorption tail is most probably due to the segregated wurtzite-type with $x\approx 0.09$ proven with PL for $x=0.35$ and $x=0.55$ in our as grown thin films. However the uncertainty in the determination of the band gap that could give rise to the low-energy absorption tail due to the strong overlapping and the presence of defects that cannot be disregarded when phase separation starts prevent us to conclude what the origin of the low-energy absorption tail is.

\citet{sans2004} showed that the pressure coefficient of the band gap of wurtzite-type Mg$_{x}$Zn$_{1-x}$O increases with $x$ and the transition pressure $P_T$ at which the wurtzite phase transforms into the rock-salt phase decreases with $x$. Therefore, if we study the band gap of the wurtzite Mg$_{0.3}$Zn$_{0.7}$O sample by optical absorption spectroscopy and PL under high pressure we would be able to i) determine the pressure coefficient of the low-energy absorption tail and ii) isolate the segregated phase which having a lower Mg content would persist in the wurtzite phase when the wurtzite phase with $x\approx 0.3$ transforms to rock-salt. This would allow us to confirm in as grown Mg$_{0.3}$Zn$_{0.7}$O the origin of the low-energy absorption tail which should present a d$E_g$/d$P \approx 25$ meV/GPa \citep{sans2004} if due to a segregated wurtzite phase with $x\approx 0.09$ and unveil the PL from the segregated phase that should be present as a consequence of the phase separation that exists for this Mg concentration [Fig. \ref{fig:fig1} (c)].

The optical absorption spectra of the as grown Mg$_{0.3}$Zn$_{0.7}$O thin film are shown at different pressures in Fig. \ref{fig:fig3} (a). Up to 7.3 GPa the shape of the absorption edge, including the low-energy tail, is kept while the absorption edge shifts to higher energies due to the volume contraction. This indicates that both the main absorption edge and the tail have a similar pressure dependence confirming that the origin of the low-energy absorption tail is due to a wurtzite phase with lower Mg content. At 7.8 GPa the absorbance of the main absorption edge decreases as a consequence of the onset of the wurtzite to rock-salt phase transition while the absorbance of the low-energy tail persists up to 10.3 GPa when only the absorption tail of the rock-salt band gap is observed indicating the end of the phase transition. The similar pressure dependence of the low-energy tail and the main absorption edge confirm that the origin of the absorption tail is related to the band gap absorption of a wurtzite phase with lower Mg content. However, the strong overlapping does not allow us to reliably quantify its band gap and thus estimate its Mg content. The energy derivative of the absorption spectra can provide an estimation of the relative proportions of both wurtzites with different Mg content and the pressure dependencies of their band gaps. In Fig. \ref{fig:fig3} (c) we show a collection of d$\alpha$/d$E$ spectra at different pressures. At 0.3 GPa two Gaussian peaks can be clearly observed at $\sim$3.5 and $\sim$4 eV. These peaks in the derivative spectrum correspond to the inflection point of the absorption edge, which roughly occurs at the band gap minus the width of the electronic transition. With this reservation, we can reliably assign the derivative peaks to the absorption edges \citep{lopez2015} for $x = 0.09$ and $x = 0.3$ supporting our previous conclusion that the low-energy tail is due to the absorption edge of the segregated phase with $x \approx 0.09$. Under pressure, the peak due to $x\approx 0.3$ blueshifts faster than the peak due to $x\approx 0.09$ up to 7.8 GPa when the intensity of the high-energy peak drops until becoming comparable to the intensity of the low-energy peak which remains constant up to 9.3 GPa. This indicates that wurtzite with $x\approx 0.3$ starts transforming to rock-salt at around 7.8 GPa while the segregated wurtzite phase remains unaffected up to 9.3 GPa at least.

\begin{figure}
\centering
\includegraphics[width=0.45\textwidth]{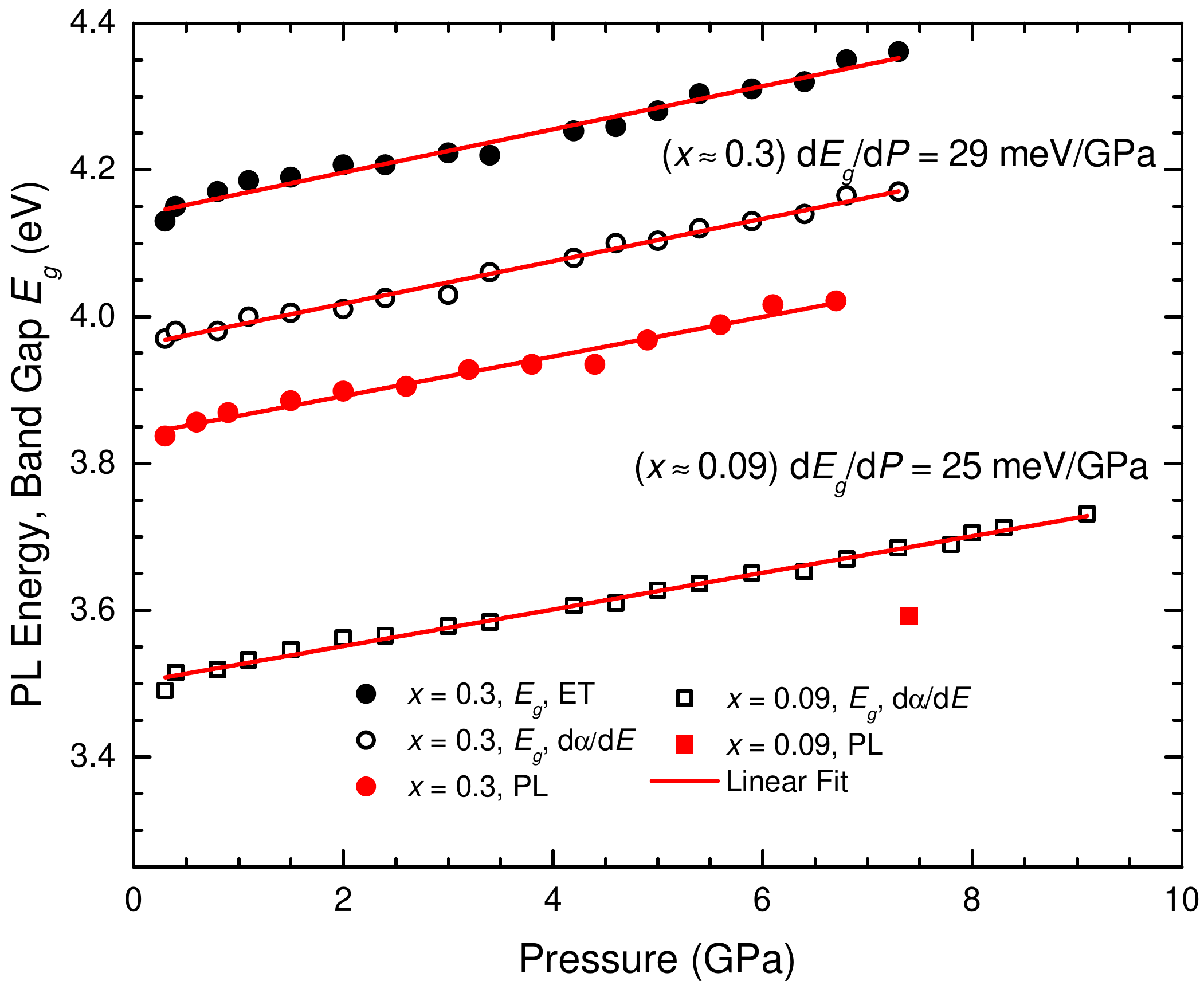}
\caption{\label{fig:fig4} Pressure dependence of the band gap $E_g$ (black) and PL peak (red) of Mg$_{0.3}$Zn$_{0.7}$O. Circles are from the wurtzite with higher Mg content and squares are from the segregated phase. The red square is the PL energy of the segregated phase ($x \approx 0.09$) only visible once wurtzite Mg$_{0.3}$Zn$_{0.7}$O has transformed to rock-salt. The solid symbols represent the band gap obtained by Elliot-Toyozawa theory and the empty symbols are the band gap obtained by the energy derivative of the absorption spectra. Red continuous lines are fits to the data points obtaining a d$E_g$/d$P$ of 25 meV/GPa for $x\approx 0.09$ in good agreement to \citep{sans2004} and 29 meV for $x\approx 0.3$.}
\end{figure}

In Fig. \ref{fig:fig2} we have found that while the PL signal from the segregated wurtzite phase is clearly observed for samples with $x=0.35$ and $x=0.55$, in the case of the sample with $x=0.3$ there is no detectable PL signal from the segregated phase. With the optical absorption experiment performed on the as grown Mg$_{0.3}$Zn$_{0.7}$O sample we have demonstrated the existence of a segregated phase with a concentration of $x \approx 0.09$. All this indicates that the amount of segregated wurtzite phase in the as grown Mg$_{0.3}$Zn$_{0.7}$O sample, though visible with optical absorption, cannot be detected with PL. The reason why this occurs might be that the PL signal of the segregated phase is too weak and appears to be masked by the signal of the dominant peak corresponding to the wurzite phase with $x\approx 0.3$. If this is the case, at around 7.3 GPa when, according to the optical absorption study the wurzite phase with $x\approx 0.3$ starts to transform into the rock-salt phase, the signal from the segregated wurtzite phase should emerge. This is what can be seen in Fig. \ref{fig:fig3} (d). The PL peak of wurzite with $x\approx 0.3$ blueshifts with pressure up to 7.4 GPa when the phase transition occurs and the intensity of the main PL peak drops. At this pressure a weak PL peak with an energy of 3.59 eV emerges. At 9.3 GPa the PL peak from the wurzite with $x\approx 0.3$ vanishes while the PL peak at 3.59 eV stays up to 10.3 GPa when the phase transitions of both wurtzite phases with different Mg concentrations have finished and the rock-salt phase shows no PL signal. According to \citet{sans2004} wurtzite Mg$_{0.09}$Zn$_{0.91}$O thin film has a pressure coefficent of 25 meV/GPa. Considering that our segregated wurtzite phase has a PL peak at 7.4 GPa of 3.59 eV, we can extrapolate an energy at ambient pressure to 3.4 eV. That is exactly the energy of the PL peak due to the segregated phase found for $x=0.35$ and $x=0.55$ at ambient conditions, and it corresponds to a sample with $x=0.09$ [Fig. \ref{fig:fig2} (a)].

Finally, in Fig. \ref{fig:fig4} we show the pressure dependence of both the band gap and the PL peak in the wurtzite phase of Mg$_{0.3}$Zn$_{0.7}$O. The observed shift of $\sim 0.3$ eV between the band gap energy and the PL peak indicates that the origin of the PL peak is not intrinsic or excitonic and can be due to certain compositional disorder. Under pressure the shift remains constant with pressure with both techniques providing a pressure coefficient for the band gap of d$E_g$/d$P$ = 29 meV/GPa ($x = 0.3$) and 25 meV/GPa ($x=0.09$). This value extends the dependence on $x$ of d$E_g$/d$P$ from $x = 0.13$ \citep{sans2004} to $x = 0.3$ and shows that d$E_g$/d$P$ saturates for high magnesium contents.
%A value smaller than the estimated 36.6 meV/GPa by \citet{sans2004} for $x=0.3$ indicating that the effect that pressure has on the band gap of wurtzite Mg$_{x}$Zn$_{1-x}$O saturates for high Mg contents.

\section{Conclusions}
In conclusion, we have shown with a spectroscopic high-pressure approach that phase segregation can be probed in as grown thin films of phase separated Mg$_{0.3}$Zn$_{0.7}$O even for small embedded volumes not detected by x-ray diffraction and not accessible by transmission electron microscopy except with annealed samples \citep{gries2015}. We have solved the controversy about the low-energy absorption tail usually observed overlapping with the main absorption edge for $x>0.3$. We have found that it is due to the band gap of the segregated wurtzite phase with $x \approx 0.09$ and not to the tail of the coexisting rock-salt phase \citep{lopez2015}. The present work shows the usefulness of high pressure optical studies to obtain relevant information about phase separation effects in semiconductor alloys.

%Finally, the phase transition pressures $P_T$ of Mg$_{0.3}$Zn$_{0.7}$O and Mg$_{0.09}$Zn$_{0.91}$O have been set to 7.4 and 9.3 GPa, which are higher values than expected by the empirical law found in Ref. \citep{sans2004}; 3.5 and 7.7 GPa. This is probably the result of using mica \citep{faust1994} instead of ScAlMgO$_4$ which is more incompressible \citep{errandonea2011} and indicates that the empirical law is no longer linear for high Mg concentrations.  

\section*{Acknowledgements}
V.M.-B and J.R.-F. thank the Universitat de Val\`encia and the Spanish MINECO for the Atracci\'o de talent and Juan de la Cierva (IJCI-2014-20513) Programs, respectively. This paper was supported by Spanish MINECO under grants MAT2016-75586-C4-1/3-P and TEC2014-60173, and by Generalitat Valenciana under projects Prometeo II 2015/004 and ISIC/2012/008.

%\bibliography{apl}

\end{document}